# Condition for Energy Efficient Watermarking with Random Vector Model without WSS Assumption

Bin Yan*, *Member, IEEE,* Zheming Lu, *Senior Member, IEEE,* and Yinjing Guo

**Abstract**

Energy efficient watermarking preserves the watermark energy after linear attack as much as possible. We consider in this letter non-stationary signal models and derive conditions for energy efficient watermarking under random vector model without WSS assumption. We find that the covariance matrix of the energy efficient watermark should be proportional to host covariance matrix to best resist the optimal linear removal attacks. In WSS process our result reduces to the well known power spectrum condition. Intuitive geometric interpretation of the results are also discussed which in turn also provide more simpler proof of the main results.

Bin Yan and Yinjing Guo are with the Department of Communication Engineering, School of Information and Electrical Engineering, Shandong University of Science and Technology. 266510, Qingdao, P. R. China. yanbin@ieee.org

Zheming Lu is with the department of Electronic and Communication Engineering, Sun Yat-sen University. 510275, GuangZhou, P. R. China. zheminglu@ieee.org

Bin Yan is the corresponding author, email: yanbinhit@hotmail.com.

# Condition for Energy Efficient Watermarking with Random Vector Model without WSS Assumption

## I. INTRODUCTION

An important question in digital watermarking is how to design the watermarks so as to best resist the attacks. In a series of papers, Su *et al.* [1] considered linear correlation detection and modeled the attack as optimal linear shift invariant filtering plus additive noise. They concluded that the best watermark that can resist the attack must have a similar power spectrum density (PSD) as that of the original host signal. In [2], Hwang *et al.* considered optimal Neyman-Pearson detection for correlated host signal model and wiener filtering attack and derived the requirement on watermark PSD. These theoretical analysis has provided the watermark practitioners with more insight into the structure of optimal watermarking system in the presence of attacks. However, their results are stated under the assumption of wide sense stationary (WSS) and ergodic random processes and in terms of PSD. In practical implementation, we sometime need a more convenient form in the language of vector matrix operations. In addition, the WSS assumption is not always satisfied by real signals such as image and speech. Question arises as whether or not the power spectrum condition is still valid for the more general random vector signal model? This work aims to answer this question and serves as complement to our previous work in [3], where we have designed a framework to shape the power spectrum of the watermark embedded by dither modulation. We derive in this paper the condition on watermark covariance matrix required for energy efficient watermarking without the WSS assumption. In section II, we discuss the optimal attack strategy of the attacker. We then formulate and solve the watermark design problem in section III. The geometric interpretation of the attack and defense are presented in section IV. Connection to WSS results is discussed in section V and finally we conclude this paper in section VI. All random vectors in this paper are assumed to be zero mean. Vectors are represented by lower case letters and matrices in upper case letters. We made no distinction between random quantities and deterministic quantities since they can be distinguished from the context. Statistical expectation is denoted by $\mathbb{E}\left(\cdot\right)$. We use $\mathbb{R}$ to represent the set of real numbers.

## II. OPTIMAL LINEAR ATTACK

### A. Matrix Wiener Filter

Since the important role played by the matrix Wiener filter in this discussion, we review it briefly. Let's consider the estimation of signal embedded in additive noise using linear filters. The signal $\mathbf{s}$ with covariance matrix $\mathbf{C_s}$ is contaminated by additive noise $\mathbf{n}$ with covariance matrix $\mathbf{C_n}$ that is independent of the signal. We observe the resulting signal $\mathbf{x} = \mathbf{s} + \mathbf{n}$ and would like to design a linear estimate $\hat{\mathbf{s}} = \mathbf{W}\mathbf{x}$ that minimize the mean squared error (MSE) between the estimated signal and the original signal. The MSE is defined as $\text{MSE} = \mathbb{E}(\hat{\mathbf{s}} - \mathbf{s})$. With straightforward matrix derivative calculation we obtain the linear minimum MSE (LMMSE) estimator as $\hat{\mathbf{s}}_{\text{MMSE}} = \mathbf{W}_{\text{MMSE}}\mathbf{x}$, where the best estimator is $\mathbf{W}_{\text{MMSE}} = \mathbf{C_s}(\mathbf{C_s} + \mathbf{C_n})^{-1}$. The error signal $\mathbf{e} = \hat{\mathbf{s}} - \mathbf{s}$ is zero mean with covariance matrix $\mathbf{C_e} = (\mathbf{I} - \mathbf{W})\mathbf{C_s}$. Implementation issues can be found in literatures like [4].

### B. Watermarking Model

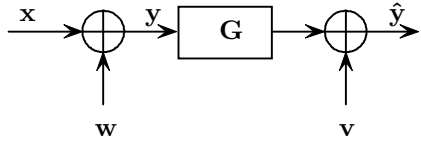

Fig. 1. Structure of embedding and attacking process.

The host signal $\mathbf{x}$ is modeled as random vector with zero mean and covariance matrix $\mathbf{C_x}$. The watermarks $\mathbf{w}$ are taken from a library of watermarks statistically characterized by zero mean and covariance matrix $\mathbf{C_w}$. We note that the statistics of the host and watermark may change over time, hence the more appropriate notation should be $\mathbf{C_x}(k)$ and $\mathbf{C_w}(k)$ with $k$ indicating the time index. But for clarity, we just drop the time index, or we just concentrate on a certain time instant. The watermark and the host signal are assumed to be independent of each other. This is true for most spread spectrum based watermarking. This assumption is also approximately true for dither modulation based watermarking if the quantization step is small enough [3]. The watermarked signal $\mathbf{y}$ is obtained by $\mathbf{y} = \mathbf{x} + \mathbf{w}$ with covariance matrix that can be found to be $\mathbf{C_y} = \mathbf{C_x} + \mathbf{C_w}$ from the independence assumption of $\mathbf{x}$ and $\mathbf{w}$. To measure the distortion between watermarked host signal $\mathbf{y}$ and original host signal $\mathbf{x}$, we use the average power $D(\mathbf{y}, \mathbf{x}) = \frac{1}{N}\mathbb{E}\left\{\|\mathbf{y} - \mathbf{x}\|^2\right\}$. At the receiving end, the watermark detection problem is actually the detection of known signal in correlated noise. A suboptimal yet convenient detection scheme is based on linear correlation $s = \frac{1}{N}\mathbf{y}^T\mathbf{w}$. As a measure of detection performance for all watermarks, it is more appropriate to average the linear correlation statistics statistically over all watermarks, so we use $r = \mathbb{E}(s)$.

### C. The Attacker's Problem

The goal of the attacker is to make the detection of watermark at the decoder invalid while keeping the distortion as small as possible [1]. We may formulate this goal as a constrained optimization problem and we call it *the attacker's problem*: Given $\mathbf{C_x}$, $\mathbf{C_w}$ and $r_0$, the attacker tries to find the best linear attack $\mathbf{G}$ and covariance matrix $\mathbf{C_v}$ of the noise $\mathbf{v}$ so as to minimize the distortion between the attacked signal and the original signal $D(\hat{\mathbf{y}}, \mathbf{x})$, subject to the constrain that $r = r_0$. To solve this problem, we need to express the distortion $D(\hat{\mathbf{y}}, \mathbf{x})$ and the average linear correlation in terms of the decision variable $\mathbf{G}$ and $\mathbf{C_v}$ of the attacker.

*1) Attack Distortion:* The attack distortion can be expressed as

$$D(\hat{\mathbf{y}}, \mathbf{x}) = \frac{1}{N}\mathbb{E}\left(\|\hat{\mathbf{y}} - \mathbf{x}\|^2\right) \tag{1}$$

$$= \frac{1}{N}\text{tr}\left(\mathbb{E}\left((\hat{\mathbf{y}} - \mathbf{x})(\hat{\mathbf{y}} - \mathbf{x})^T\right)\right) \tag{2}$$

$$= \frac{1}{N}\left[\text{tr}\left(\mathbb{E}\left(\hat{\mathbf{y}}\hat{\mathbf{y}}^T\right) - \mathbb{E}\left(\hat{\mathbf{y}}\mathbf{x}^T\right) - \mathbb{E}\left(\mathbf{x}\hat{\mathbf{y}}^T\right) + \mathbb{E}\left(\mathbf{x}\mathbf{x}^T\right)\right)\right]. \tag{3}$$

Since $\hat{\mathbf{y}}$ is the host signal after linear attacks, we have $\hat{\mathbf{y}} = \mathbf{Gy} + \mathbf{v} = \mathbf{G}(\mathbf{x} + \mathbf{w}) + \mathbf{v}$. Finally we get the expression for attack distortion as

$$D(\hat{\mathbf{y}}, \mathbf{x}) = \frac{1}{N}\left[\text{tr}\left(\mathbf{GC_xG}^T\right) + \text{tr}\left(\mathbf{GC_wG}^T\right) - \text{tr}\left(\mathbf{GC_x}\right) - \text{tr}\left(\mathbf{G}^T\mathbf{C_x}\right) + \text{tr}\left(\mathbf{C_v}\right) + \text{tr}\left(\mathbf{C_x}\right)\right] \tag{4}$$

*2) Average Linear Correlation:* After averaged statistically over all possible watermarks, the average linear correlation can be expressed as

$$r = \frac{1}{N}\mathbb{E}\left(\hat{\mathbf{y}}^T\mathbf{w}\right) = \frac{1}{N}\text{tr}\left(\mathbb{E}\left(\hat{\mathbf{y}}\mathbf{w}^T\right)\right) = \frac{1}{N}\text{tr}\left(\mathbf{GC_w}\right) \tag{5}$$

### D. Solving the Attacker's Problem

With the expression for the attack distortion and averaged detection statistics, the attacker's problem is equivalent to the following constrained optimization problem: Choose $\mathbf{G}$ and $\mathbf{C_v}$ to

$$\min \quad \frac{1}{N}\left[\text{tr}\left(\mathbf{GC_xG}^T + \mathbf{GC_wG}^T - \mathbf{GC_x} - \mathbf{G}^T\mathbf{C_x} + \mathbf{C_v} + \mathbf{C_x}\right)\right] \tag{6}$$

$$\text{s.t.} \quad \frac{1}{N}\text{tr}\left(\mathbf{GC_w}\right) = r_0 \tag{7}$$

To solve this, we introduce the Lagrangian as $\mathcal{L} = D + \lambda(r - r_0)$. Differenting formally of $\mathcal{L}$ w.r.t. the matrix $\mathbf{G}$ and setting the result to zero we get

$$\frac{\partial \mathcal{L}}{\partial \mathbf{G}} = \mathbf{G}\mathbf{C_x}^T + \mathbf{G}\mathbf{C_x} + \mathbf{G}\mathbf{C_w}^T + \mathbf{G}\mathbf{C_w} - \mathbf{C_x}^T - \mathbf{C_x} + \lambda \mathbf{C_w}^T = 0$$

So the optimal linear attack is

$$\mathbf{G} = \left(\mathbf{C_x} - \frac{\lambda}{2}\mathbf{C_w}\right)(\mathbf{C_x} + \mathbf{C_w})^{-1}.$$

$D$ is linear in $\operatorname{tr}(\mathbf{C_v})$ and $r$ is independent of $\mathbf{C_v}$, so to minimize $D$ w.r.t. $\mathbf{C_v}$, we choose $\operatorname{tr}(\mathbf{C_v})$ to be its minimum possible value. Since $\operatorname{tr}(\mathbf{C_v}) \geq 0$, so the minimum $\operatorname{tr}(\mathbf{C_v})$ is zero. This in turn implies that the variance of the components of $\mathbf{v}$ are all zeros, which implies that $\mathbf{v} = 0$ since $\mathbf{v}$ has zero mean. So the best attack should introduce no additional additive noise that is independent of the host and watermark. It is instructive to write the linear attack in terms of the matrix Wiener filter we discussed above:

$$\begin{align}
\mathbf{G} &= \left(\mathbf{C_x} + \mathbf{C_w} - \mathbf{C_w} - \frac{\lambda}{2}\mathbf{C_w}\right)(\mathbf{C_x} + \mathbf{C_w})^{-1} \tag{8}\\
&= \mathbf{I} - \left(1 + \frac{\lambda}{2}\right)\mathbf{C_w}(\mathbf{C_x} + \mathbf{C_w})^{-1} \tag{9}\\
&= \mathbf{I} - \gamma \mathbf{H}, \tag{10}
\end{align}$$

where $\gamma = 1 + \frac{\lambda}{2}$ and $\mathbf{H} = \mathbf{C_w}(\mathbf{C_x} + \mathbf{C_w})^{-1}$ is the Wiener filter that estimates the watermark from the watermarked host signal $\mathbf{y}$.

### III. THE WATERMARK DESIGN PROBLEM

The watermark designer is aware of the existence of the optimal linear attack as discussed in last section. He has no control on this attack, but he can design his watermark to resist the attack as much as possible [1]. The watermark designer wishes to maximize the residual watermark energy after attack by designing the covariance matrix of the watermark appropriately, subject to the constrain on perceptual quality. We refer to this as the *watermark design problem*. We study below how the the residual watermark energy and perceptual constrain are related to watermark covariance.

*A. The Residual Watermark Power*

The optimal attack $\mathbf{G} = \mathbf{I} - \gamma \mathbf{H}$ can be considered as first estimate watermark $\hat{\mathbf{w}}$ from the watermarked signal $\mathbf{y}$ by $\hat{\mathbf{w}} = \mathbf{H}\mathbf{y} = \mathbf{H}(\mathbf{x} + \mathbf{w})$ and then subtract $\gamma \mathbf{w}$ from the watermarked signal. We concentrate on the special case of removal attack where $\gamma = 1$. It is easy to show that $\hat{\mathbf{w}}$ is zero mean with covariance matrix

$$\mathbf{C}_{\hat{\mathbf{w}}} = \mathbb{E}\left(\hat{\mathbf{w}}\hat{\mathbf{w}}^T\right) \tag{11}$$

$$= \mathbb{E}\left(\mathbf{H}\left(\mathbf{x}+\mathbf{w}\right)\left(\mathbf{x}^T+\mathbf{w}^T\right)\mathbf{H}^T\right) \tag{12}$$

$$= \mathbf{H}\mathbf{C}_{\mathbf{x}}\mathbf{H}^T + \mathbf{H}\mathbf{C}_{\mathbf{w}}\mathbf{H}^T. \tag{13}$$

So the average energy in the estimated watermark is

$$\mathbb{E}\left(\hat{\mathbf{w}}^T\hat{\mathbf{w}}\right) = \operatorname{tr}\left(\mathbf{H}\mathbf{C}_{\mathbf{x}}\mathbf{H}^T + \mathbf{H}\mathbf{C}_{\mathbf{w}}\mathbf{H}^T\right). \tag{14}$$

It is not difficult to show that $\mathbb{E}\left(\hat{\mathbf{w}}^T\mathbf{w}\right) = \mathbb{E}\left(\hat{\mathbf{w}}^T\hat{\mathbf{w}}\right).$

The amount of average watermark power left in the host after the attack is related to the decision variable $\mathbf{C}_{\mathbf{w}}$ by:

$$E = \frac{1}{N}\mathbb{E}\left(\|\mathbf{w}-\hat{\mathbf{w}}\|^2\right) \tag{15}$$

$$= \frac{1}{N}\mathbb{E}\left(\operatorname{tr}\left((\mathbf{w}-\hat{\mathbf{w}})(\mathbf{w}-\hat{\mathbf{w}})^T\right)\right) \tag{16}$$

$$= \frac{1}{N}\left[\operatorname{tr}\left(\mathbf{C}_{\mathbf{w}}\right) - \operatorname{tr}\left(\mathbf{C}_{\hat{\mathbf{w}}}\right)\right] \tag{17}$$

$$= \frac{1}{N}\left[\operatorname{tr}\left(\mathbf{C}_{\mathbf{w}}\right) - \operatorname{tr}\left(\mathbf{C}_{\mathbf{w}}\left(\mathbf{C}_{\mathbf{x}}+\mathbf{C}_{\mathbf{w}}\right)^{-1}\mathbf{C}_{\mathbf{w}}\right)\right]. \tag{18}$$

The perceptual difference between the original host signal and the watermarked host signal is measured by the average power of the difference between the two signals. Suppose that the embedding induced distortion is constrained to be $\bar{P}_w$. We then obtain the following restriction on the covariance matrix of the watermark $\frac{1}{N}\mathbb{E}\left(\|\mathbf{w}\|^2\right) = \frac{1}{N}\operatorname{tr}\left(\mathbf{C}_{\mathbf{w}}\right) = \bar{P}_w$.

*B. Solving the Watermark Design Problem*

Summarizing the above results, we have formulated the watermark design problem as: Choose an appropriate $\mathbf{C}_{\mathbf{w}}$ to

$$\max \quad E = \frac{1}{N}\operatorname{tr}\left(\mathbf{C}_{\mathbf{w}} - \mathbf{C}_{\mathbf{w}}\left(\mathbf{C}_{\mathbf{x}}+\mathbf{C}_{\mathbf{w}}\right)^{-1}\mathbf{C}_{\mathbf{w}}\right) \tag{19}$$

$$\text{s.t.} \quad \frac{1}{N}\operatorname{tr}\left(\mathbf{C}_{\mathbf{w}}\right) = \bar{P}_w \tag{20}$$

To solve this problem, we need two important facts from matrix derivative. The detailed discussion of matrix differentiation can be found in [5]. Suppose that $\mathbf{U}, \mathbf{V}$ and $\mathbf{Y}$ below are matrices of appropriate dimension. They are all functions of the matrix argument $\mathbf{X}$, then we have the following two facts:

$$\text{Fact1}: \quad \frac{\partial}{\partial \mathbf{X}}\operatorname{tr}\left(\mathbf{U}\mathbf{V}\right) = \frac{\partial}{\partial \mathbf{X}}\operatorname{tr}\left(\mathbf{U}_c\mathbf{V}\right) + \frac{\partial}{\partial \mathbf{X}}\operatorname{tr}\left(\mathbf{U}\mathbf{V}_c\right). \tag{21}$$

January 11, 2009 DRAFT5

The terms with subscript $c$ are considered as constants during differentiation.

$$\text{Fact2}: \quad \frac{\partial}{\partial \mathbf{X}}\text{tr}\left(\mathbf{Y}^{-1}\right) = -\frac{\partial}{\partial \mathbf{X}}\text{tr}\left(\mathbf{Y}_c^{-2}\mathbf{Y}\right). \quad (22)$$

We construct the Lagrangian of the watermark design optimization problem as

$$\mathcal{L}\left(\mathbf{C_x}, \lambda\right) = E + \lambda\left[\frac{1}{N}\text{tr}\left(\mathbf{C_w}\right) - \bar{P}_w\right]. \quad (23)$$

To find the stationary point, we need

$$\frac{\partial \mathcal{L}}{\partial \mathbf{C_w}} = \frac{1}{N}\frac{\partial}{\partial \mathbf{C_w}}\text{tr}\left(\mathbf{C_w}\right) - \frac{1}{N}\frac{\partial}{\partial \mathbf{C_w}}\text{tr}\left(\mathbf{C_w}\left(\mathbf{C_w} + \mathbf{C_x}\right)^{-1}\mathbf{C_w}\right) + \frac{\lambda}{N}\frac{\partial}{\partial \mathbf{C_w}}\text{tr}\left(\mathbf{C_w}\right). \quad (24)$$

For the first term in (24), we have

$$\frac{1}{N}\frac{\partial}{\partial \mathbf{C_w}}\text{tr}\left(\mathbf{C_w}\right) = \frac{1}{N}\mathbf{I}. \quad (25)$$

To find the second term in (24), remember the property of trace $\text{tr}\left(\mathbf{ABC}\right) = \text{tr}\left(\mathbf{BCA}\right)$. So the second term in (24) is equivalent to

$$\frac{1}{N}\frac{\partial}{\partial \mathbf{C_w}}\text{tr}\left(\underbrace{\left(\mathbf{C_w} + \mathbf{C_x}\right)^{-1}}_{\mathbf{U}}\underbrace{\mathbf{C_w}^2}_{\mathbf{V}}\right) = \frac{1}{N}\left[\underbrace{\frac{\partial}{\partial \mathbf{C_w}}\text{tr}\left(\mathbf{U}_c\mathbf{C_w}^2\right)}_{\text{I}} + \underbrace{\frac{\partial}{\partial \mathbf{C_w}}\text{tr}\left(\left(\mathbf{C_w} + \mathbf{C_x}\right)^{-1}\mathbf{V}_c\right)}_{\text{II}}\right] \quad (26)$$

$$, \quad (27)$$

where we have applied Fact 1 stated above. Term I in this equation is a second order term of $\mathbf{C_w}$, so we have

$$\text{I} = \left(\mathbf{C_w}\mathbf{U}_c + \mathbf{U}_c\mathbf{C_w}\right)^T = \mathbf{C_w}\left(\mathbf{C_x} + \mathbf{C_w}\right)^{-1} + \left(\mathbf{C_x} + \mathbf{C_w}\right)^{-1}\mathbf{C_w}, \quad (28)$$

since the matrices involved here are all symmetric. For term II, if we define $\mathbf{Y}^{-1} = \left(\mathbf{C_w} + \mathbf{C_x}\right)^{-1}\mathbf{V}_c$ and apply Fact 2 stated above, we then obtain

$$\text{II} = -\frac{\partial}{\partial \mathbf{C_w}}\text{tr}\left(\mathbf{Y}_c^{-2}\mathbf{Y}\right) \quad (29)$$

$$= -\frac{\partial}{\partial \mathbf{C_w}}\text{tr}\left(\mathbf{Y}_c^{-2}\mathbf{V}_c^{-1}\left(\mathbf{C_x} + \mathbf{C_w}\right)\right) \quad (30)$$

$$= -\left(\mathbf{Y}_c^{-2}\mathbf{V}_c^{-1}\right)^T \quad (31)$$

$$= -\left[\left(\mathbf{C_x} + \mathbf{C_w}\right)^{-1}\mathbf{C_w}^2\left(\mathbf{C_x} + \mathbf{C_w}\right)^{-1}\right]. \quad (32)$$

Where we have again used the symmetric property of the matrix $\mathbf{C_x}$ and $\mathbf{C_w}$ and the fact that $\frac{\partial}{\partial \mathbf{X}}\text{tr}\left(\mathbf{BX}\right) = \mathbf{B}^T$. Summarizing the above result, the stationary point occur when

$$(1 + \lambda)\mathbf{I} + \left(\mathbf{C_x} + \mathbf{C_w}\right)^{-1}\mathbf{C_w}^2\left(\mathbf{C_x} + \mathbf{C_w}\right)^{-1} - \mathbf{C_w}\left(\mathbf{C_x} + \mathbf{C_w}\right)^{-1} - \left(\mathbf{C_x} + \mathbf{C_w}\right)^{-1}\mathbf{C_w} = 0. \quad (33)$$

After some simple matrix algebra, we have

$$\mathbf{C_w} = \frac{1-\sqrt{-\lambda}}{\sqrt{-\lambda}}\mathbf{C_x} \text{ or } \mathbf{C_w} = \frac{1+\sqrt{-\lambda}}{-\sqrt{-\lambda}}\mathbf{C_x}. \quad (34)$$

Since the diagonal elements of both $\mathbf{C_x}$ and $\mathbf{C_w}$ are positive, we discard the second case. Substitute the first case into the power constrain, we have

$$\frac{1-\sqrt{-\lambda}}{\sqrt{-\lambda}} = \frac{\bar{P}_w}{\bar{P}_x}, \quad (35)$$

where $\bar{P}_x = \frac{1}{N}\text{tr}(\mathbf{C_x})$ is the average power in the signal vector $\mathbf{x}$. Substituting (35) into (34), we then find the covariance matrix of the best watermark

$$\mathbf{C_w} = \frac{\bar{P}_w}{\bar{P}_x}\mathbf{C_x}. \quad (36)$$

Eq.(36) says that the watermark that best resist the Wiener removal attack should have a covariance matrix that is scaled version of the covariance matrix of the host signal.

## IV. Geometric Interpretation

One advantage of the result in vector-matrix form is the intuitive geometric interpretation. If we define the inner product between two random vectors $\mathbf{u}$ and $\mathbf{v}$ as $<\mathbf{u},\mathbf{v}> = \mathbb{E}\left(\mathbf{u}^T\mathbf{v}\right)$, and assume that all random vectors have bounded second order moments, then we get a complete Hilbert space $\mathcal{H}$ [6]. The Wiener removal attack discussed above can be illustrated as in Fig. 2. The subspace spanned by $\mathbf{y}$ is $\mathcal{U}_\mathbf{y} = \{\mathbf{Gy} : \mathbf{G} \in \mathbb{R}^{m\times n}\}$. Here we have chosen the horizontal plan to represent $\mathcal{U}_\mathbf{y}$. The attacker actually first project the watermark onto the subspace $\mathcal{U}_\mathbf{y}$ to get the estimated watermark $\hat{\mathbf{w}}$, he then subtract from the received signal $\mathbf{y}$ the estimated watermark to get the attacked signal $\hat{\mathbf{y}}$.

The watermark designer is interested in reducing the length of $\hat{\mathbf{w}}$. So under what condition will the length of $\hat{\mathbf{w}}$ be reduced to it's minimum? We summarize the answer to this question as a theorem stated below.

*Theorem 1:* Let $\mathcal{H}$ be the Hilbert space of random vectors with inner product defined by $\langle\mathbf{u},\mathbf{v}\rangle = \mathbb{E}\left(\mathbf{u}^T\mathbf{v}\right)$. $\mathbf{x}$ and $\mathbf{w}$ are elements of this Hilbert space. $\mathbf{x} \perp \mathbf{w}$ and $\mathbf{y} = \mathbf{x} + \mathbf{w}$. The subspace spanned by $\mathbf{y}$ is $\mathcal{U}_\mathbf{y} = \{\mathbf{Gy} : \mathbf{G} \in \mathbb{R}^{n\times n}\}$. Let $\hat{\mathbf{w}}$ be the orthogonal projection of $\mathbf{w}$ on $\mathcal{U}_\mathbf{y}$. Then $\|\hat{\mathbf{w}}\|$ is minimized when $\hat{\mathbf{w}} = \alpha\mathbf{y}$ for some $0 < \alpha < 1$.

*Proof:* Referring to Fig. 2, let $\mathbf{u}$ be the orthogonal projection of $\mathbf{w}$ on $\mathbf{y}$ and $\hat{\mathbf{w}}$ be the orthogonal projection of $\mathbf{w}$ on $\mathcal{U}_\mathbf{y}$. From Pythagorean theorem in Hilbert space, we know that

$$\|\mathbf{w}\|^2 = \|\mathbf{u}\|^2 + \|\mathbf{w}-\mathbf{u}\|^2 = \|\hat{\mathbf{w}}\|^2 + \|\mathbf{w}-\hat{\mathbf{w}}\|^2. \quad (37)$$

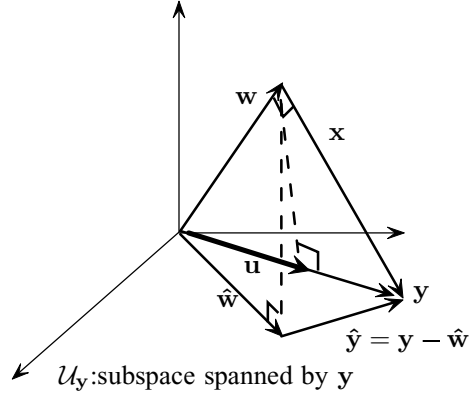

$\mathcal{U}_\mathbf{y}$:subspace spanned by $\mathbf{y}$

Fig. 2. Geometric interpretation of Wiener removal attack

From projection theorem we have $\|\mathbf{w} - \mathbf{u}\|^2 \leq \|\mathbf{w} - \hat{\mathbf{w}}\|^2$. So we have $\|\mathbf{u}\|^2 \leq \|\hat{\mathbf{w}}\|^2$, with equality if the projection $\hat{\mathbf{w}}$ is aligned with $\mathbf{y}$, namely when $\hat{\mathbf{w}} = \alpha \mathbf{y}$ for some positive $\alpha < 1$. ∎

Now The watermark design problem in Hilbert space can be stated as follows: Fix the length of the watermark in $\mathcal{H}$, i.e., $\frac{1}{N}\text{tr}(\mathbf{C_w}) = \bar{P}_x$, design the covariance matrix $\mathbf{C_w}$ of the watermark to minimize the projection of $\mathbf{w}$ on $\mathcal{U}_\mathbf{y}$. To solve the design problem, we know from Theorem 1 that the minimum projection occurs when $\hat{\mathbf{w}} = \alpha \mathbf{y}$. This is equivalent to $\mathbf{Hy} = \alpha \mathbf{y}$. Combining this with the fact that the Wiener filter is $\mathbf{H} = \mathbf{C_w}(\mathbf{C_x} + \mathbf{C_w})^{-1}$, we obtain $\mathbf{C_w}(\mathbf{C_x} + \mathbf{C_w})^{-1} = \alpha$. After some algebra we get $\mathbf{C_w} = \frac{\alpha}{1-\alpha}\mathbf{C_x}$. This is equivalent to say that the watermark covariance matrix should be proportional to the host covariance matrix. To find the undetermined coefficient $\alpha$, we substitute above formula into the power constrain of the watermark to obtain $\frac{\alpha}{1-\alpha} = \frac{\bar{P}_w}{\bar{P}_x}$. So finally we obtain the same results as in (36) above. We note that with geometric interpretation, we can abandon the tedious formal differentiation.

## V. Connection to WSS Case

We show in this section that under the WSS assumption, the result stated in (36) reduces to the results obtained in [1]. As we know, the covariance matrices $\mathbf{C_x}$ and $\mathbf{C_w}$ are all symmetric Toeplitz matrices. For Toeplitz matrix $\mathbf{C_x}$, as $N \to \infty$, its eigenvalues and corresponding eigenvectors approach

$$\lambda_{x,i} = \Phi_{ww}(f_i) \tag{38}$$

$$\mathbf{v}_{x,i} = \frac{1}{\sqrt{N}}\begin{bmatrix} 1 & e^{j2\pi f_i} & \cdots & e^{j2\pi(N-1)f_i} \end{bmatrix}^T, \tag{39}$$

where $f_i = \frac{i}{N}$ and $\Phi_{xx}(f)$ is the PSD of the host signal. For watermark covariance matrix $\mathbf{C_w}$, we obtain similar results with eigenvalues $\lambda_{w,i}$ and the corresponding eigenvectors $\mathbf{v}_{x,i}$. So we can express

(36) as

$$\sum_{i=0}^{N-1} \lambda_{w,i} \mathbf{v}_{w,i} \mathbf{v}_{w,i}^T = \sum_{i=0}^{N-1} \frac{\sigma_w^2}{\sigma_x^2} \lambda_{x,i} \mathbf{v}_{x,i} \mathbf{v}_{x,i}^T, \qquad (40)$$

where we have used the fact that for WSS signal, $\bar{P}_w = \sigma_w^2$ and $\bar{P}_x = \sigma_x^2$. So for WSS host and watermark, as $N \to \infty$, we get the exactly the same result as stated in [1]:

$$\Phi_{ww}(f) = \frac{\sigma_w^2}{\sigma_x^2} \Phi_{xx}(f). \qquad (41)$$

## VI. CONCLUSION

We have derived in this paper the condition for energy efficient watermarking without resorting to the WSS assumption. The analysis result says that, for energy efficient watermarking, the covariance matrix of the watermark ensemble should be proportional to the covariance matrix of the host ensemble. The power spectrum condition can be directly derived from our results by applying the WSS condition. The geometric interpretation of the attacks and defenses in Hilbert space of random vectors provides us intuitive interpretation and simpler proof.